# An equivalent-effect phenomenon in eddy current non-destructive testing of thin structures

**Wuliang Yin[1,2],(Senior Member, IEEE), Jiawei Tang[2], Mingyang Lu[2,*], Hanyang Xu[2], Ruochen Huang[2], Qian Zhao[3], Zhijie Zhang[1], Anthony Peyton[2]**
[1]School of Instrument and Electronics, North University of China, Taiyuan, Shanxi, 030051 China
[2]School of Electrical and Electronic Engineering, University of Manchester, Manchester, M13 9PL UK
[3]College of Engineering, Qufu Normal University, Shandong, 273165 China

*Corresponding author: Mingyang Lu (mingyang.lu@manchester.ac.uk).

**ABSTRACT** The inductance/impedance due to thin metallic structures in non-destructive testing (NDT) is difficult to evaluate. In particular, in Finite Element Method (FEM) eddy current simulation, an extremely fine mesh is required to accurately simulate skin effects especially at high frequencies, and this could cause an extremely large total mesh for the whole problem, i.e. including, for example, other surrounding structures and excitation sources like coils. Consequently, intensive computation requirements are needed. In this paper, an equivalent-effect phenomenon is found, which has revealed that alternative structures can produce the same effect on the sensor response, i.e. mutual impedance/inductance of coupled coils if a relationship (reciprocal relationship) between the electrical conductivity and the thickness of the structure is observed. By using this relationship, the mutual inductance/impedance can be calculated from the equivalent structures with much fewer mesh elements, which can significantly save the computation time. In eddy current NDT, coils inductance/impedance is normally used as a critical parameter for various industrial applications, such as flaw detection, coating and microstructure sensing. Theoretical derivation, measurements and simulations have been presented to verify the feasibility of the proposed phenomenon.

## I. INTRODUCTION

Previously, massive works have been proposed on the electromagnetic eddy current evaluation techniques. And basic simulation methods can be summarized as Method of Auxiliary Sources (MAS), Boundary-Element Method (BEM), and Finite-Element Method (FEM) [1]-[5]. The fundamental principle of MAS is introducing a source within and near the surface of the structure that can scatter the same electromagnetic field as that around the structure. Then each step of the electromagnetic field change can be equivalent to a change or addition of a new source. The merit of MAS is simplifying the eddy currents computation procedure, as therefore, increasing the efficiency of the calculation. BEM is essentially only modelling or meshing the boundary region of the structure, which can significantly reduce the elements and calculation amount. However, both MAS and BEM are not commonly used or even cannot evaluate the eddy currents of the structure with sophisticated geometry. For instance, the MAS method is hard to compute the eddy currents of the structure with the rough surface; and BEM cannot do the eddy current computation of structures with non-linear geometry. Although FEM needs to mesh/model the whole structure or even the space surrounded by the structure (considering the eddy current skin/diffusion effect), it is the most widely used and can solve eddy current evaluations for almost all types of structures including non-linear geometry and material properties.

For the FEM applied in the electromagnetic area, considerable research works have been published on the eddy current testing theory under low frequency or even the static electromagnetic field. However, little has been discussed on the high-frequency eddy current computation especially for the metallic structure with high conductivity (HC), which will encounter some computation issues especially the eddy current skin/diffusion effects [6]. As more intensive induced eddy currents are distributed in the structure surface underneath the sensor under the eddy current skin/diffusion effect, significantly refining the mesh around the surface region especially underneath the sensor area is necessary to maintain the simulation accuracy when using the FEM method. However, models/meshes with intensive elements will result in a mass of computation burden.

The impedance calculation of thin plates or even thin shell under high-frequency is still a challenge for the researchers and industrial engineers in Non-destructive Testing (NDT) area such as crack detection and material property inspection [7]. For the crack detection, the variations of the measured thickness can be a criterion for the defect detection when moving the sensor along the plate. For example, a small sensor can be used to detect the extent of the surface damage for the metallic pipe. For the canonical FEM formulas carried out by Bíró [8], an A-V edge element formulation was proposed and proved to be better on mitigating the high-frequency eddy current simulation problem due to the skin/diffusion effect. The advantage of the A-V formulation on alleviating the skin/diffusion effects is reducing the singularity of the stiffness matrix corresponding to the structure model. Others techniques exist in [9] and [10]. However, most of those works still require extensive calculations as the structure mesh/model remains the same; and the solver still needs to compute the whole elements. Essentially, reducing the rank of the structure system stiffness matrix is imperative in relieving the computation burden caused by the skin/diffusion effect.

In this paper, based on the transverse electric (TE) propagation algorithms through medium with different material layers, an equivalent-effect phenomenon is discovered, in which a reciprocal relationship is found between the electrical conductivity and thickness of the thin tested piece for the same sensor output signal response - impedance/inductance. The fundamental principle of the proposed equivalent-effect phenomenon is using an alternative thicker structure but with less conductivity to have almost the same impedance value as the original structure. With the proposed equivalent-effect phenomenon, the impedance computation burden will be significantly reduced, which can be explained by two aspects. Firstly, under the same frequency, a conductive metallic structure with much lower conductivity will be less affected by the diffusion/skin effect. Consequently, fewer elements are needed for meshing/modelling the less conductive structure. Secondly, the thinner structure requires finer element size in order to stay at the same accuracy. As a result, much intensive and more overall mesh elements are needed for the eddy current computation of the original structure's each layer. Most important of all, the measured signal - mutual impedance is almost immune to the altering of the structure's electrical and geometric properties controlled by the found equivalent-effect phenomenon. The detected sensor response signal - mutual impedance at the sensor's terminals is usually treated as the basic parameter for the flaw inspection especially in the non-destructive testing/evaluation (NDT/NDE) applications [11]-[16]. Overall, based on the proposed equivalent-effect phenomenon, an alternative thicker structure with less conductivity can be equivalent to the thin metallic layer modelling, which can significantly reduce mesh size without affecting the detected mutual impedance.

## II. THEORETICAL BASE FOR THE EQUIVALENT-EFFECT PHENOMENON

### A. TRANSMITTER-RECEIVER MUTUAL INDUCTANCE

Intuitively, in order to get the same inductance value, we might think the relation between the thickness and electrical conductivity of two different samples should be the same as that between the skin depth and electrical field. However, the sensor detected signal is not solely determined by the eddy current (density) within the samples, which is controlled by the electrical conductivity; the detected signal also depends on the attenuation distance of the electromagnetic waves during the propagation, which is related to the thickness of the sample. Therefore, it is necessary to do a step-by-step analyzation of the decay during the propagation (including transmissions and reflections) of the wave field. Currently, there are three wave modes to investigate the propagation of the electromagnetic wave - Transverse Electric and Magnetic (TEM) mode, Transverse Electric (TE) mode, and Transverse Magnetic (TM) mode. In this paper we have utilized TE wave mode to analyze the induced voltages between the two coupling coils of the sensor.

The transmitter-receiver mutual inductance changes caused by the tested piece can be obtained by applying the equation presented by Dodd and Deeds [17]. For this article, a generalized equation of calculating induced voltage that could be applied to any coil sensor is present:

$$V = j\omega \int_v \mathbf{A} \cdot ds = -\int_v \mathbf{E} \cdot ds = 2\pi r E_r \quad (1)$$

Here, $E_r$ denotes magnitude of the electrical field on the coils position; $v$ indicates the region of the sensor coil; r is the radius of the sensor coil.

Then, the mutual inductance of the tested specimen and sensor system should be,

$$\Delta L = \frac{2\pi r E_r}{j\omega I} \quad (2)$$

Where, I denotes the amplitude of the excitation current.

### B. SOLUTIONS OF ELECTRICAL FIELD TERM $E_r$ – TRANSVERSE ELECTRIC(TE) WAVE MODE

The following derivations are associated with the solution of electrical field term $E_r$ using transverse electric (TE) wave propagation algorithms through medium with different material layers.

Since the sensor coils are parallel to the sample slab, the majority of the electrical field excited should parallel to the surface of the planar sample. Consequently, the electrical field in the planar-layered media can be calculated via the Transverse electric (TE) wave algorithms. According to the equations presented by Chow in book <Waves and Fields in Inhomogeneous Media> [18], the reflection and transmission of TE wave in a half-space (Fig. 1) obeys the following equations. It is necessary to mention that the y-axis in Fig.1 is perpendicular to x-z plane and positive towards inside.

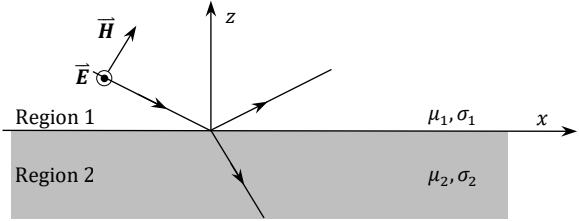

FIGURE 1. Reflection and transmission of a plane wave at an interface

Consider both the incident and reflected waves are presented, the electrical field of upper half-space can be written as,

$$E_{1y}(z) = E_0 e^{k_1 z} + E_0 R_{12} e^{-k_1 z} \quad (3)$$

Where, $R_{12}$ is the ratio of reflected wave amplitude to the incident wave amplitude; $E_0$ is the electrical field prior to propagate (i.e. the electrical field on the sensor position when the sensor is put in free space). $z$ denotes the depth related to the excitation coil.

In the lower half-space, however, only a transmitted wave is present; hence a general expression is,

$$E_{2y}(z) = E_0 T_{12} e^{k_2 z} \quad (4)$$

Where, $T_{12}$ is the ratio of transmitted wave amplitude to the incident wave amplitude.

In equation (3) and (4), $k_i$ is a constant which equals,

$$k_i = \sqrt{\alpha_0^2 + j\omega\sigma_i\mu_i} \quad (5)$$
$$i = 1, 2 \text{ (Region number)} \quad (6)$$

Where, $\sigma_i$ and $\mu_i$ indicate the electrical conductivity and magnetic permeability of upper half-space; $\alpha_0$ is a spatial frequency constant, which is solely affected by the sensor. $\alpha_0$ is defined to be 1 over the smallest dimension of the coil [22].

In equation (3) and (4), $R_{ij}$ and $T_{ij}$ are the Fresnel reflection and transmission coefficients from region $i$ to region $j$,

$$R_{ij} = \frac{\mu_j k_i - \mu_i k_j}{\mu_j k_i + \mu_i k_j} \quad (7)$$

$$T_{ij} = \frac{2\mu_j k_i}{\mu_j k_i + \mu_i k_j} \quad (8)$$

Further, for a three-layer medium, a series of the reflection and transmission coefficients during the TE wave propagation are shown in Fig.2. Normally, multiple reflections occurred in region 2. Since the TE wave would decay significantly after several reflections, here only the TE waves prior to the third reflection within region 2 are analysed.

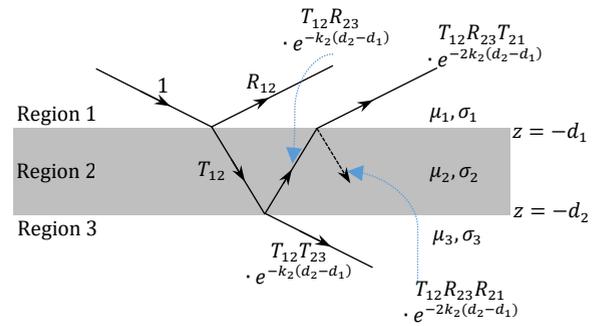

FIGURE 2. Geometric series of TE wave reflection and transmission in a three-layer medium

Therefore, for region 1, the wave can be written as,

$$E_{1y}(z) = E_0\big(e^{k_1 z} + \tilde{R}_{12} e^{-(z+2d_1)k_1}\big) \quad (9)$$

Where, $\tilde{R}_{12}$ is the ratio of the amplitude for all the reflected waves in region 1 to the amplitude of original incident wave.

Similarly, the wave in region 2 is,

$$E_{2y}(z) = E_1\big(e^{k_2 z} + R_{23} e^{-(z+2d_1)k_2}\big) \quad (10)$$

Where, $R_{23}$ is the Fresnel reflection coefficient for a down-going wave in region 2 reflected by region 3.

The wave in region 3 can be written as,

$$E_{3y}(z) = E_2 e^{k_3 z} \quad (11)$$

In region 2, the down-going wave is caused by the transmission of the down-going wave in region 1 plus a reflection of the up-going wave in region 2. Consequently, at the interface $z = -d_1$, the constraint condition obeys,

$$E_1 e^{-k_2 d_1} = E_0 T_{12} e^{-k_1 d_1} + E_1 R_{21} R_{23} e^{-2k_2(2d_2 - d_1)} \quad (12)$$

From equation (12), $E_1$ can be obtained in terms of $E_0$, yielding,

$$E_1 = \frac{T_{12} R_{23} T_{21} e^{-2k_2(d_2 - d_1)}}{1 - R_{21} R_{23} e^{-2k_2(d_2 - d_1)}} \quad (13)$$

In region 1, the overall reflected wave is a consequence of the reflection of the down-going wave in region 1 plus a transmission of the up-going wave in region 2. Consequently, at the interface $z = -d_1$, the constraint condition obeys,

$$E_0 \tilde{R}_{12} e^{-k_1 d_1} = E_0 R_{12} e^{-k_1 d_1} + E_1 T_{21} R_{23} e^{-2k_2(2d_2 - d_1)} \quad (14)$$

Substitute (13) into (14), it can be derived,

$$\tilde{R}_{12} = R_{12} + \frac{T_{12} R_{23} T_{21} e^{-2k_2(d_2 - d_1)}}{1 - R_{21} R_{23} e^{-2k_2(d_2 - d_1)}} \quad (15)$$

Substitute (15) into (9), the electrical wave present in region 1 can be obtained.

$$E_{1y}(z) = E_0 \left(e^{k_1 z} + \left(R_{12} \frac{T_{12} R_{23} T_{21} e^{-2k_2(d_2 - d_1)}}{1 - R_{21} R_{23} e^{-2k_2(d_2 - d_1)}}\right) e^{-(z+2d_1)k_1}\right) \quad (16)$$

For a planar non-magnetic sample with an electrical conductivity of $\sigma_0$ and a small thickness of $D_0$, the electrical field on the position of the sensor becomes,

$$E_{1y}(0) = E_0 \left(1 + \frac{k_1 - k_2}{k_1 + k_2}\left(1 - \frac{4k_1 k_2 e^{-2k_2 D_0}}{(k_1 + k_2)(k_1 + k_2) - (k_2 - k_1)(k_2 - k_1)e^{-2k_2 D_0}}\right)e^{-2D_0 k_1}\right) \quad (17)$$

Where, $k_1 = a_0, k_2 = \sqrt{a_0^2 + j\omega\sigma_0\mu_0} \quad (18)$

Further deduction from equation (17)

$$E_{1y}(0) = E_0 \left(1 + \left(\frac{(k_2-k_1)(k_2+k_1)-(k_2-k_1)(k_2+k_1)e^{2k_2 D_0}}{-(k_2-k_1)(k_2-k_1)+(k_2+k_1)(k_2+k_1)e^{2k_2 D_0}}\right)e^{-2D_0 k_1}\right) \quad (19)$$

For the thin structure, with $k_1 a_0 \ll 1$ holds, the term $e^{-2D_0 k_1}$ can be substituted with $1 - 2D_0 k_1 \approx 1$. Then, equation (19) can be approximated as,

$$E_{1y}(0) = E_0 \left(1 + \frac{(k_2-k_1)(k_2+k_1)-(k_2-k_1)(k_2+k_1)e^{2k_2 D_0}}{-(k_2-k_1)(k_2-k_1)+(k_2+k_1)(k_2+k_1)e^{2k_2 D_0}}\right) \quad (20)$$

Therefore, for the equation (2), the electrical field on the coils position should be,

$$E_r = E_0 \left(1 + \frac{(k_2-k_1)(k_2+k_1)-(k_2-k_1)(k_2+k_1)e^{2k_2 D_0}}{-(k_2-k_1)(k_2-k_1)+(k_2+k_1)(k_2+k_1)e^{2k_2 D_0}}\right) \quad (21)$$

Substituting equation (21) to equation (2), tested sensor-sample mutual inductance is

$$L = \frac{2\pi r}{j\omega I}\left|\left(1 + \frac{(k_2-k_1)(k_2+k_1)-(k_2-k_1)(k_2+k_1)e^{2k_2 D_0}}{-(k_2-k_1)(k_2-k_1)+(k_2+k_1)(k_2+k_1)e^{2k_2 D_0}}\right)E_0\right|$$

$$= \frac{2\pi r}{j\omega I}\left(1 + \frac{(k_2-k_1)(k_2+k_1)-(k_2-k_1)(k_2+k_1)e^{2k_2 D_0}}{-(k_2-k_1)(k_2-k_1)+(k_2+k_1)(k_2+k_1)e^{2k_2 D_0}}\right)E_0 \quad (22)$$

Similarly, when the tested region is free space, the tested inductance is,

$$L_{air} = \frac{2\pi r E_0}{j\omega I} \quad (23)$$

Combining (22) with (23), the mutual inductance of sensor-sample system is,

$$\Delta L = L - L_{air} = \frac{2\pi r E_0}{j\omega I} \frac{(k_2-k_1)(k_2+k_1)-(k_2-k_1)(k_2+k_1)e^{2k_2 D_0}}{-(k_2-k_1)(k_2-k_1)+(k_2+k_1)(k_2+k_1)e^{2k_2 D_0}} \quad (24)$$

Where, $k_1 = a_0$, $k_2 = \sqrt{a_0^2 + j\omega\sigma_0\mu_0}$, $E_0$ is the magnitude of the electrical field on the sensor position when the sensor is put in free space.

Substituting $e^{-2k_2 D_0}$ with $1 - 2k_2 D_0$,

$$\Delta L = \frac{2\pi r E_0}{j\omega I}\frac{-j\omega\sigma_0\mu_0 D_0}{a_0 + 2a_0^2 D_0 + jD_0\omega\sigma_0\mu_0 - 2D_0 a_0 D_0\sqrt{a_0^2 + j\omega\sigma_0\mu_0}} \quad (25)$$

For the thin structure, with $D_0 a_0 \ll 1$ holds, equation (25) can be approximated as,

$$\Delta L = \frac{2\pi r E_0}{I}\frac{-\mu_0 D_0 \sigma_0}{a_0 + j\omega\mu_0 D_0 \sigma_0} \quad (26)$$

It can be found from equation (26) that, for a specific operation frequency $\omega$, an excitation current $I$ and same sensor setup ($k_1$ or $a_0$ is a constant), the transmitter-receiver inductance $\Delta L$ is solely controlled by the sample via the assembled term $D_0 \sigma_0$.

Therefore, for two thin structures with different electrical conductivities ($\sigma_1$ and $\sigma_2$) and thickness ($D_1$ and $D_2$), their inductance are found to be nearly identical $\Delta L(\sigma) = \Delta L(D)$ only if

$$D_1/D_2 = \sigma_2/\sigma_1 \quad (27)$$

## III. VALIDATION METHODS

### A. VALIDATION SETUP

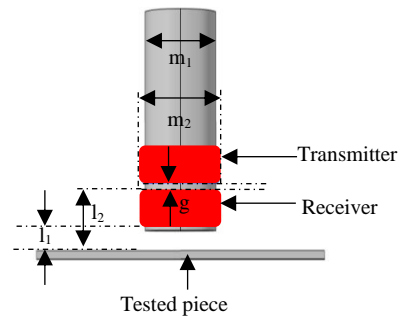

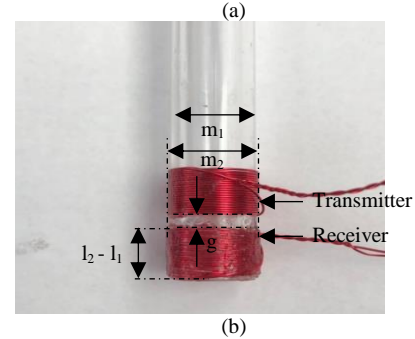

**FIGURE 3.** Air-core T-R sensor (a) simulation setup (b) experimental setup

TABLE I
SENSOR PARAMETERS

| Parameter | Value |
|---|---|
| Inner diameter $m_1$ (mm) | 12 |
| Outer diameter $m_2$ (mm) | 12.63 |
| Coils height $l_2-l_1$ (mm) | 8 |
| Coils gap g (mm) | 2 |
| Lift-offs $l_1$ (mm) | 1 |
| Coils turns $N_1/N_2$ | 25/25 |

In this paper, the metallic sample is tested by an air-cored probe with two coupled coils attached. As shown in Fig. 3, the top coil and bottom coil are the sensor's transmitter and receiver. An alternating excitation current with a range of operation frequency flows in the transmitter, which is used to produce the EM wave. Moreover, the sample's inductance can be derived from the induced voltage detected by the receiver. The parameters of the sensor are illustrated in Table I.

In this paper, both the edge-element FEM and analytical solution are used to validate the proposed equivalent-effect phenomenon.

### B. VALIDATION METHOD – EDGE-ELEMENT FEM

For the edge-element FEM solver, a software package based on the presented FEM solver has been built, which uses the Bi-conjugate Gradients Stabilised (CGS) iterative method to solve the matrix. Compared with the canonical EM simulation solver, the novelty of this FEM software package is that it has assigned the solution for the previous frequency to be the initial guess of the next frequency. I.e. the presented FEM solver is more efficient than the conventional EM simulation solver on the multi-frequency spectra calculations [20].

## C. VALIDATION METHOD – DODD AND DEEDS ANALYTICAL SOLUTION

The Dodd Deeds analytical solution is chosen to be the analytical solution of the forward problem solver on the inductance computation of the metallic slab.

The Dodd Deeds analytical solution indicates the inductance change of air-core coupled coils caused by a layer of the metallic slab for both non-magnetic and magnetic cases [17]. The difference in the complex inductance is $\Delta L(\omega) = L(\omega) - L_A(\omega)$, where the mutual inductance between two coils above a plate is $L(\omega)$, and $L_A(\omega)$ is the inductance of coil caused by the free space.

Both the FEM and Dodd Deeds solver were scripted by MATLAB, which are computed on a ThinkStation P510 platform with a Dual Intel Xeon E5-2600 v4 Processor and 32GB RAM.

## IV. RESULTS

Firstly, an experiment has been carried out. The motivation of the experiment is to check the performance of the found phenomenon when using a real sensor. The experimental data has its noise more or less. We want to check whether the error caused by the noise is negligible for equation (27). Once equation (27) is validated by the experiment, further parts of section are all solely calculated by the analytical solution (Dodd Deeds) and FEM.

To validate the proposed equivalent-effect phenomenon as shown in equation (27), a copper plate sample with a thickness of 0.56 mm and an electrical conductivity of 59.8 MS/m is treated as the original structure as shown in Figure 4 (a) and Figure 5 (a). A corresponding brass sample with a thickness of 2.00 mm and an electrical conductivity of 16.7 MS/m is served as the equivalent structure, as shown in Figure 4 (a) and Figure 3 (b). The correlation between the electrical conductivity and thickness of these two samples obeys the equation (27). The planar dimensions of these structures are all 80×50 mm.

For the experimental setup, as shown in Figure 5 (b), Zurich impedance instrument [21] has been used to measure the air-core sensor induced signal response – mutual impedance/inductance of the sensor influenced by the tested samples. The working frequency range of the Zurich instruments is from 1 kHz to 500 kHz. The amplitude of the excitation current is 10 mA.

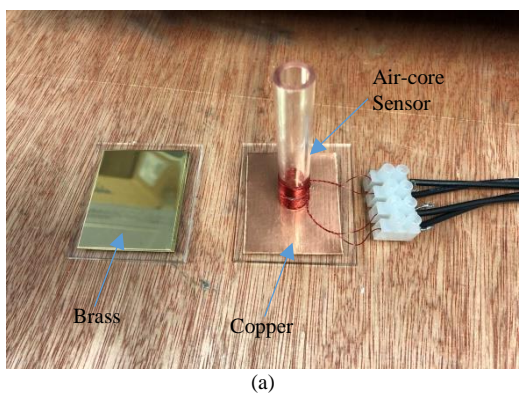

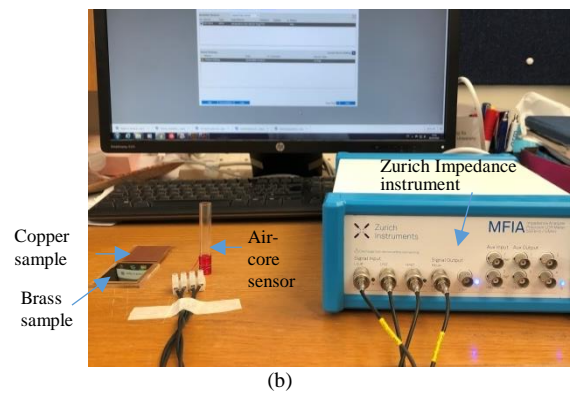

(b)

**FIGURE 4.** Experimental setup (a) the brass and copper sample (b) Zurich Impedance measurements system

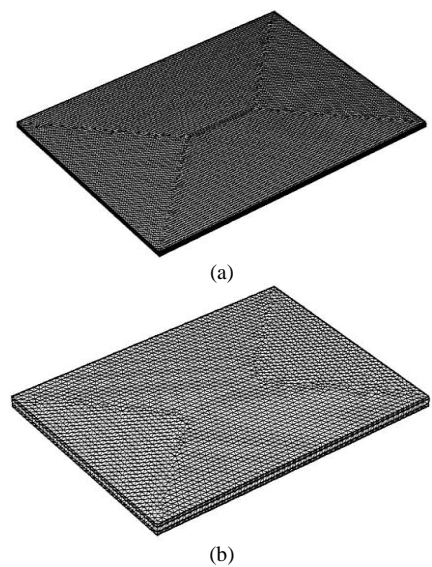

**FIGURE 5.** Mesh modelling of thicker structures a) Original copper structure b) equivalent brass structure

For the simulation modelling, both the Finite-Element method (FEM) and analytical solution (Dodd Deeds) have been used to calculate the mutual inductance. Table II and Table III illustrate the parameters and modelling element dimensions for the brass and copper samples. At the end of this part, the eddy current distributions for both structures are presented and discussed.

The lower limit of the number of the element depends on how accurate of the inductance we want to obtain. Few elements can influence the accuracy of the results. Therefore, we have set a criterion for the under limit of the element – the error of the calculated inductance via FEM is within 3% when compared to that of analytical solution. For the determination of the element, we have utilized COMSOL to generate the element meshes. In Table III, the maximum and minimum element sizes are proportional to the thickness of the structure. The maximum element growth rate represents the elements dimension maximum changing rate among the adjacent subdomains. The smaller maximum element growth rate is, the more homogeneous element sizes are. The curvature factor in Table III shows how bent the structure surface is. The larger curvature factor is the more intensive structure surface elements are. Maximum element growth rate and curvature factor are identical for structures with different thickness in order to maintain the same mesh resolution for different depth eddy current simulation.

TABLE II
MODELLING PARAMETERS OF THICKER STRUCTURES

|  | Original structure mesh modelling | Equivalent structure mesh modelling |
|---|---|---|
| Electrical conductivity (MS/m) | 59.8 | 16.7 |
| Thickness (mm) | 0.56 | 2.00 |
| The number of mesh elements | 157482 (~ 157 k) | 41702 (~ 42 k) |

TABLE III
FREE TETRAHEDRAL ELEMENT DIMENSIONS INFORMATION FOR THE THICKER STRUCTURES

|  | Original structure mesh modelling | Equivalent structure mesh modelling |
|---|---|---|
| Maximum element size (mm) | 0.10 | 0.27 |
| Minimum element size (mm) | 0.05 | 0.14 |
| Maximum element growth rate | 1.20 | 1.20 |
| Curvature factor | 0 | 0 |

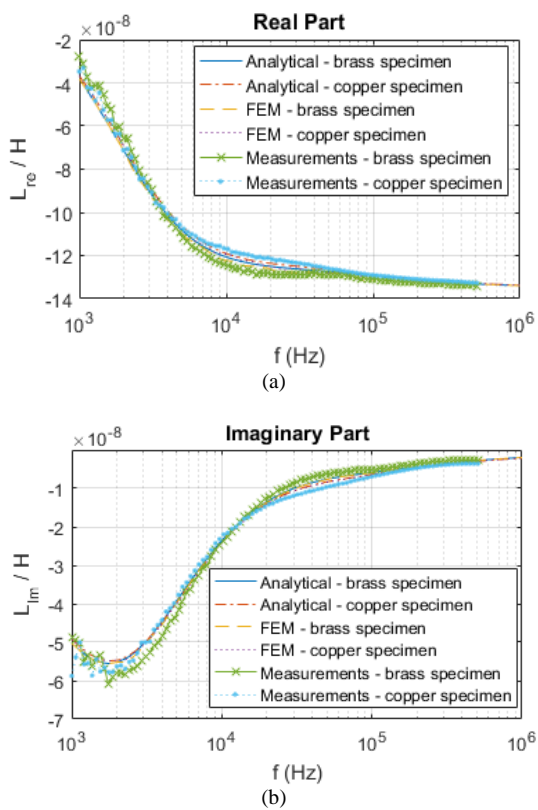

FIGURE 6. Simulations and measured results of copper and brass structures muti-frequency inductance spectra a) real part b) imaginary part

In Figure 6, it is found that the multi-frequency inductance curve for the equivalent brass structure (electrical conductivity $\sigma$ - 16.7 MS/m, thickness $D$ - 2.00 mm) can coincide well with that for the original copper structure (electrical conductivity $\sigma$ - 59.8 MS/m, thickness $D$ - 0.56 mm) especially under high frequencies (over 100 kHz). The maximum error between the original and equivalent structure inductance-frequency curve from measurements and simulations results computed by both FEM and the analytical solution is only 3.1% under the high-frequency over 100 kHz.

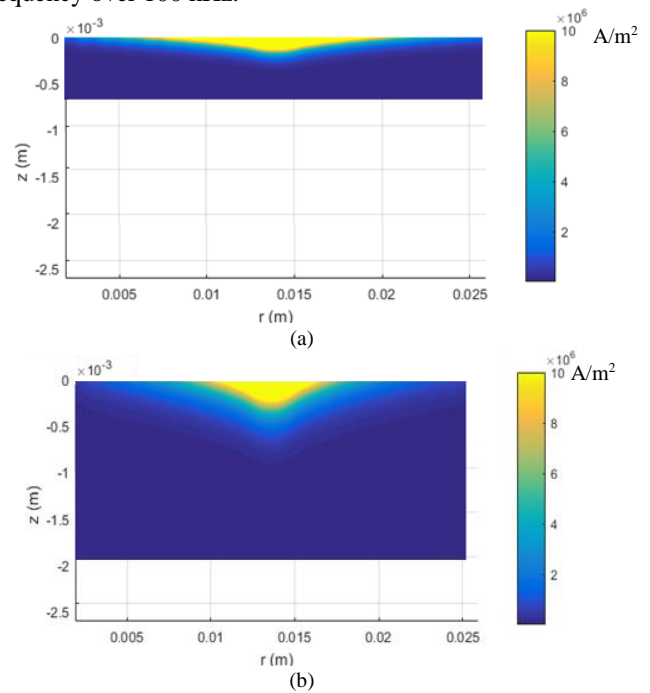

FIGURE 7. Eddy current distributions for structures with different electrical conductivities and thickness under an operation frequency of 500 kHz a) The original copper structure with an electrical conductivity of 59.8 MS/m, and thickness of 0.56 mm b) The equivalent brass structure an electrical conductivity of 16.7 MS/m, and thickness of 2.00 mm

In Figure 7, the original copper structure shows a broader and larger eddy currents padding area than that in the equivalent brass structure under the same colour bar criterion. Therefore, more intensive mesh elements are required for the computation of the original copper structure multi-frequency inductance. Further, a small thickness of the original copper structure will also result in more intensive meshed elements in order to remain the same number of depth samples for the eddy current simulations.

Although the equivalent-effect phenomenon is verified to be accurate especially under the high operating excitation frequencies, the performance of this phenomenon on thinner specimens is worth to be analysed further.

Since both the presented FEM and the analytical solution are verified to be accurate by comparing with the measured results, the following further validations only focus on the FEM and analytical solutions.

## V. APPLICABILITY OF THE EQUIVALENT-EFFECT PHENOMENON

### A. THICKNESS AND ELECTRICAL CONDUCTIVITY INFLUENCE

Considering the magnetic flux may penetrate thinner metallic plates, the fitting performance between the multi-frequency inductance curves for a thinner original structure and the corresponding equivalent structure, may differ from that of the thicker metallic plates.

The sensor detected multi-frequency inductance for different structures (structures with different electrical conductivities and thickness) including an original aluminium structure and an equivalent structure are analysed in this section.

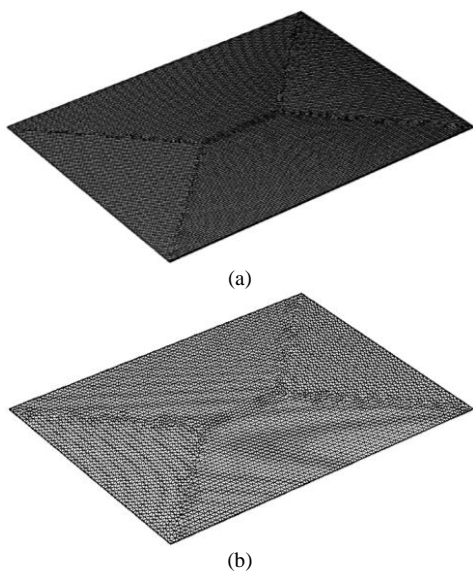

(a)

(b)

**FIGURE 8.** Mesh modelling of thin structures a) original aluminium structure b) equivalent structure

In this work, the sensor setup is shown in Figure 3. Four types of structures – an original aluminium structure, a low-conductive structure with the same thickness, a thicker structure with same electrical conductivity, and an equivalent structure are used to investigate the conductivity and thickness influences on the inductance under multi-frequency operation. The meshed structures for the original and equivalent structure are shown in Figure 8. The planar sizes of all the structures are 80 × 50 mm. The original and equivalent structures' parameters – electrical conductivity, thickness and number of mesh elements are listed in Table IV. The electrical conductivity of the equivalent structure is calculated by the proposed co-relation equation (equation (27)) between the thickness and electrical conductivity change in the equivalent structure parameters evaluation part. Table V denotes the dimensions of the free tetrahedral element for the original and equivalent structures.

TABLE IV
MODELLING PARAMETERS

|  | Original structure mesh modelling | Equivalent structure mesh modelling |
|---|---|---|
| Electrical conductivity (MS/m) | 36.9 | 13.5 |
| Thickness (µm) | 20 | 55 |
| The number of mesh elements | 289897(~ 290 k) | 63667(~ 63 k) |

TABLE V
FREE TETRAHEDRAL ELEMENT DIMENSIONS INFORMATION

|  | Original structure mesh modelling | Equivalent structure mesh modelling |
|---|---|---|
| Maximum element size (µm) | 2.00 | 5.50 |
| Minimum element size (µm) | 1.00 | 2.75 |
| Maximum element growth rate | 1.20 | 1.20 |
| Curvature factor | 0 | 0 |

In this work, the multi-frequency inductance spectra are computed by Finite-Element method (FEM) and analytical solution (Dodd Deeds). By contrasting the multi-frequency inductance curve for both the original aluminium structure (electrical conductivity $\sigma$ - 36.9 MS/m, thickness $D$ - 20 µm) and low-conductive structure with same thickness (electrical conductivity $\sigma$ – 13.5 MS/m, thickness $D$ - 20 µm) in Figure 9, it is found that reducing the electrical conductivity will result in right shift of the inductance multi-frequency curve. However, by comparing the curve for the original aluminium structure and the thicker structure with same electrical conductivity (electrical conductivity $\sigma$ - 36.9 MS/m, thickness $D$ - 55 µm), the inductance multi-frequency curve is shown with a left shift for increased thickness. Therefore, a controlled increased thickness can nearly compensate the shift of the inductance multi-frequency curve caused by a reduced electrical conductivity under almost all the frequencies from 10 to 1 MHz, as shown from multi-frequency curve for both the original aluminium structure and the equivalent structure (electrical conductivity $\sigma$ – 13.5 MS/m, thickness $D$ - 55 µm) in the Figure 9. Although the multi-frequency inductance curve for the original structure is coincident with that for the equivalent structure, the computation works for the original structure will be much more than that for the equivalent structure due to the extensive numbers of mesh elements (original structure – 290 k, about 5 times than the element number of equivalent structure – 63 k). The maximum error between the original and equivalent structure inductance-frequency curve computed by both FEM and the analytical solution is only 2.3% for all the frequencies ranged from 10 to 1 MHz.

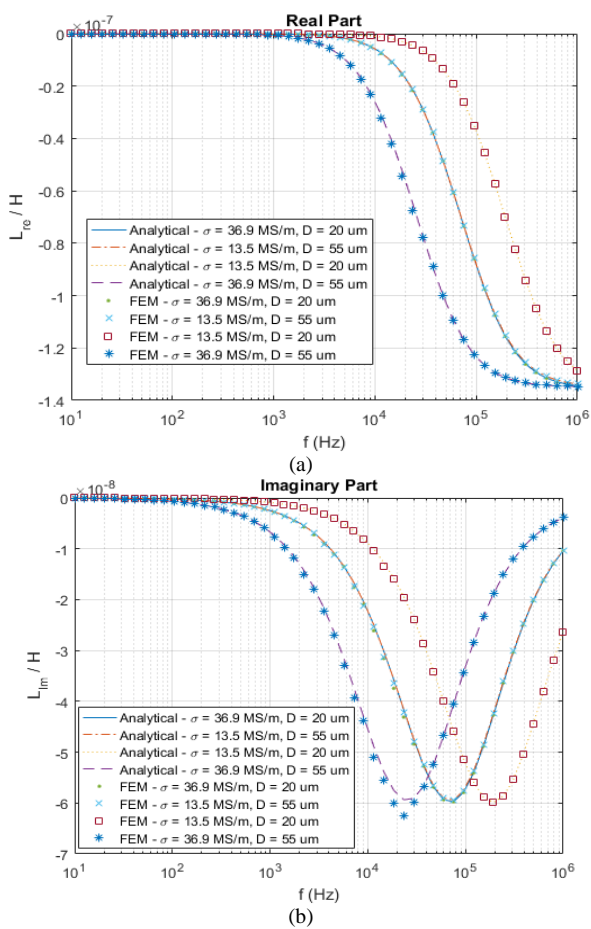

FIGURE 9. Real and imaginary parts of structures muti-frequency inductance spectra a) real part b) imaginary part

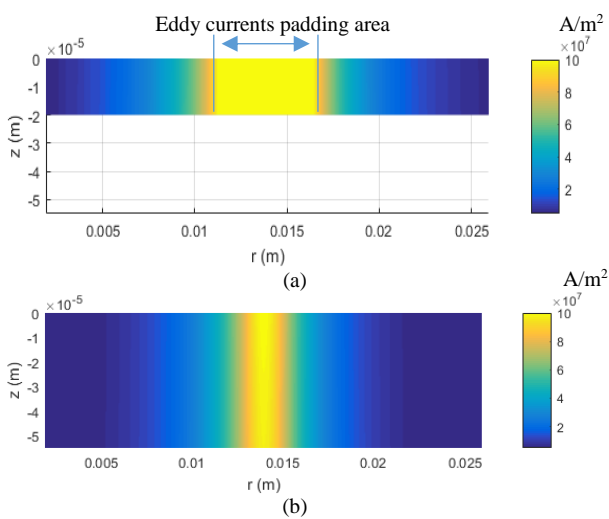

FIGURE 10. Eddy current distributions for different structures under an operation frequency of 1 MHz a) The original aluminium structure with an electrical conductivity of 36.9 MS/m, and thickness of 20 μm b) The equivalent structure an electrical conductivity of 13.5 MS/m, and thickness of 55 μm

Since the skin effect will be encountered in the original structure under the high frequencies such as 500 kHz as shown in Figure 10, a more intensive mesh is needed for the area near to the surface of the structure. As a result, the number of mesh element for the original aluminium structure (157 k) is much more significant and almost four times than that of the equivalent structure (42 k). However, much less intensive mesh elements are needed for the original aluminium structure inductance computation under low-frequency due to the reduced skin effect. In conclusion, for the thinner metallic plates, the inductance can be calculated from the original structure under nearly all the operating frequencies from 10 Hz to 1 MHz.

Even if the equivalent-effect phenomenon is valid for the flat plates geometry, its performance on the structures with other geometries such as curved plates is worth investigating, as shown in the following.

### B. INFLUENCE OF DIFFERENT CURVATURE

Compared with the results of the analytical solution, the FEM is verified to be accurate enough for the calculation of sensor-structure mutual inductance from the analysis of the equivalent-effect phenomenon performance on the flat plates as shown above. Moreover, the analytical solution can only be used to calculate the mutual inductance for the flat plate. Therefore, for the curved plate structures, the following mutual inductance is only computed by the FEM.

In this work, two types of structure modelling are used to analyze the performance of the equivalent-effect phenomenon on the curved plate's geometry structures. As shown in Figure 11 a) and c), the first one is the original aluminium curved plates mesh with a thickness of 20 μm. The structure has meshed into several layers in order to offer sufficient element samples for an accurate calculation of mutual inductance. Considering the eddy current skin/diffusion effects under high-frequency during the computation process, the empty region above and underneath the structure is also meshed. As shown in Figure 11 b) and d), the second sample is the corresponding equivalent curved plates with a thickness of 55 μm. The parameters and element dimensions for these two modelling are listed in Table VI and Table VII. Since the curved plate needs more fine mesh near to the surface of the structure, the planar dimensions for all the modelling are selected to be a smaller value of 2×2 mm. Accordingly, the diameter of the sensor for these curved modelling is chosen to be a smaller size of 0.4 mm.

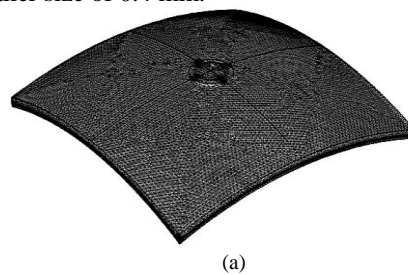

(a)

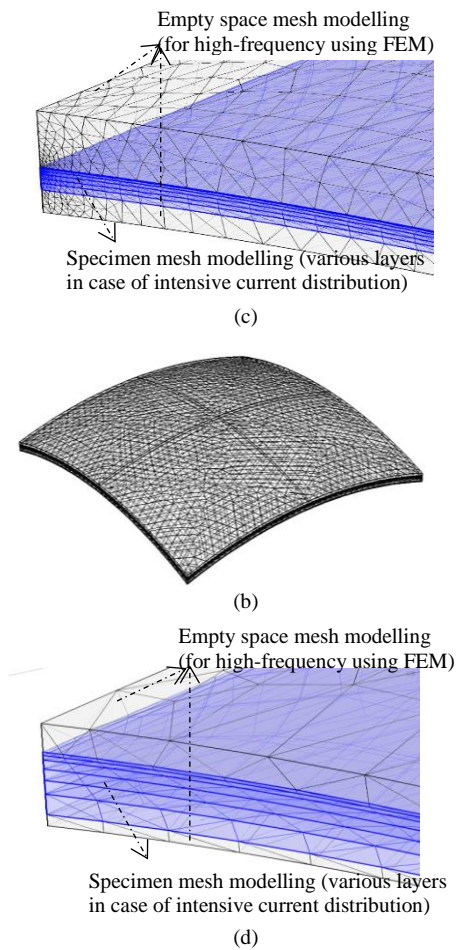

FIGURE 11. Curved structures a) Original meshed structure b) Equivalent structure c) Zooming in view of original meshed structure d) Zooming in view of the equivalent structure

TABLE VI
MODELLING PARAMETERS FOR THE CURVED STRUCTURES

|  | Original meshed structure | Equivalent structure |
|---|---|---|
| Electrical conductivity (MS/m) | 36.9 | 13.5 |
| Thickness (μm) | 20 | 55 |
| The number of mesh elements | 671480(~ 671 k) | 98230(~ 98 k) |

TABLE VII
FREE TETRAHEDRAL ELEMENT DIMENSIONS INFORMATION FOR THE CURVED STRUCTURES

|  | Original meshed structure | Equivalent structure |
|---|---|---|
| Maximum element size (μm) | 0.64 | 2.75 |
| Minimum element size (μm) | 0.30 | 1.38 |
| Maximum element growth rate | 1.50 | 1.50 |
| Curvature factor | 0.60 | 0.60 |

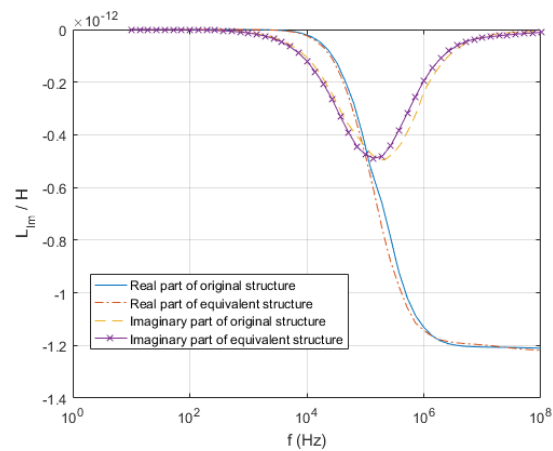

FIGURE 12. Real and imaginary parts of the curved structures mutifrequency inductance spectra

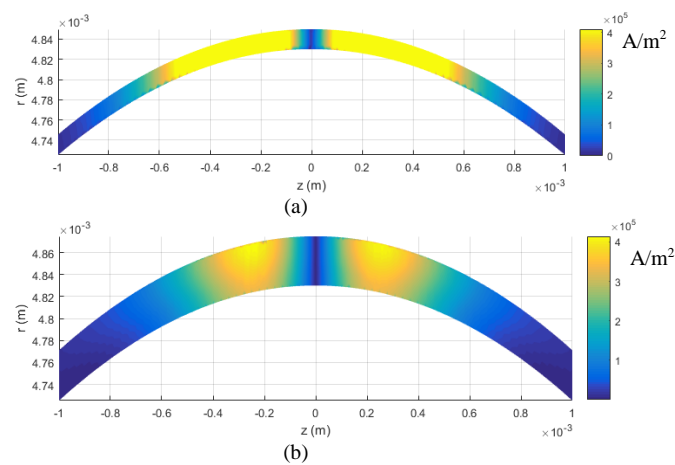

FIGURE 13. Eddy current distributions for curved structures with different electrical conductivities and thickness under an operation frequency of 1 MHz a) The original aluminium curved structure with an electrical conductivity of 36.9 MS/m, and thickness of 20 μm b) The equivalent curved structure an electrical conductivity of 13.5 MS/m, and a thickness of 55 μm

In Figure 12, the mutual inductance curve for the original structure shows a stable and well fitting (with a maximum error of 2.2%) with that for the equivalent structure. However, the original structure (671 k) needs more than six times numbers of elements than the equivalent structure (98 k), as listed in Table VI. Therefore, the equivalent-effect phenomenon shows even better performance on the curved structure inductance calculation.

It can be seen from Figure 13 that, the eddy current distributions for the original mesh modelling shows a more intensive and broader padding area than that for the equivalent structure. Hence, a more fine mesh is required for the mutual inductance calculation of the original curved structure.

### C. COMBINED EFFECTS OF DIFFERENT MATERIALS AND CURVATURE

In order to further test the feasibility of the equivalent-effect phenomenon, a copper model with an electrical conductivity of 59.8 MS/m and larger curvature is tested as follows.

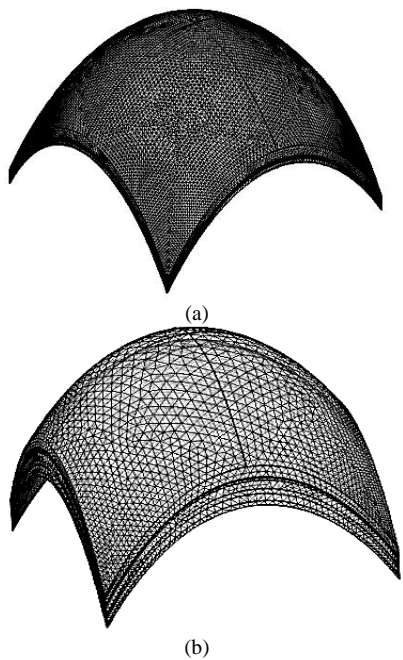

FIGURE 14. Mesh modelling of the structures with a larger curvature a) Original copper structure b) Equivalent structure

In this work, a bent copper plate with a thickness of 20 µm and a larger curvature factor of 1.20 is treated as the original structure as shown in Figure 14 (a). The thickness of the equivalent structure with an electrical conductivity of 17.3 MS/m is calculated to be approximately 69 µm by the proposed equivalent-effect phenomenon equation (equation (27)). Table VIII and Table IX illustrate the parameters and element dimensions for the thicker structures. The sensor-sample mutual inductance is computed by FEM method for both the original meshed structure and equivalent structure. Since the curved plate needs more fine meshes near to the surface of the structure, the planar dimensions for all the modelling are selected to be a smaller value of 2×2 mm. Correspondingly, the diameter of the sensor for these curved modelling is selected to be a smaller size of 0.4 mm.

TABLE VIII
MODELLING PARAMETERS FOR THE STRUCTURES WITH A LARGER CURVATURE

|  | Original meshed structure | Equivalent structure |
|---|---|---|
| Electrical conductivity (MS/m) | 59.8 | 17.3 |
| Thickness (µm) | 20 | 69 |
| The number of mesh elements | 678611 (~ 678 k) | 90268 (~ 90 k) |

TABLE IX
FREE TETRAHEDRAL ELEMENT DIMENSIONS INFORMATION FOR THE STRUCTURES WITH A LARGER CURVATURE

|  | Original meshed structure | Equivalent structure |
|---|---|---|
| Maximum element size (µm) | 1.60 | 8.66 |
| Minimum element size (µm) | 0.80 | 4.33 |
| Maximum element growth rate | 1.50 | 1.50 |
| Curvature factor | 1.20 | 1.20 |

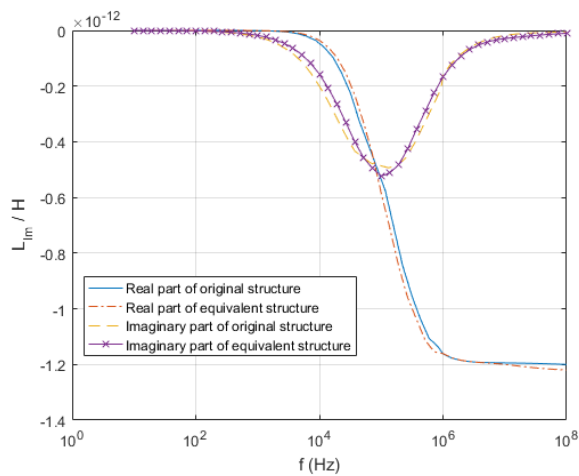

FIGURE 15. Real and imaginary parts of the multi-frequency inductance spectra for the structures with a larger curvature

It can be seen from Figure 15 that the mutual inductance curve for the original structure shows a stable and well fitting (with a maximum error of 2.7%) with that for the equivalent structure. However, the original structure (678 k) needs more than seven times numbers of elements than the equivalent structure (90 k), as listed in Table VIII.

By comparing Figure 16 (a) and Figure 16 (b), the eddy current distributions for the original mesh modelling shows a more intensive and broader padding area than that for the equivalent structure. Hence, in order to get the accurate value of the sensor-sample mutual inductance, a more fine mesh is needed.

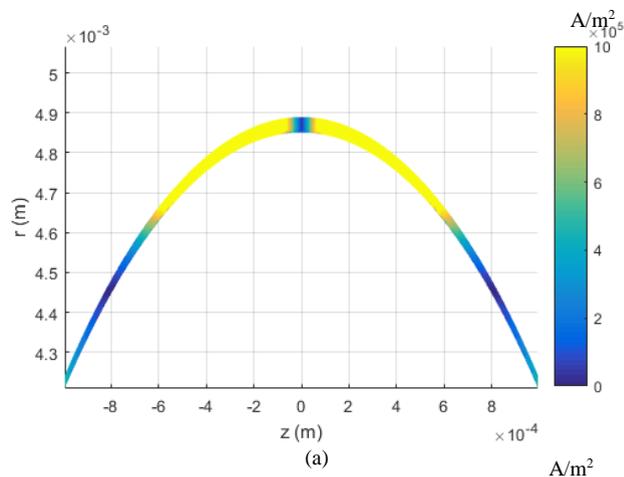

(a)

A/m²

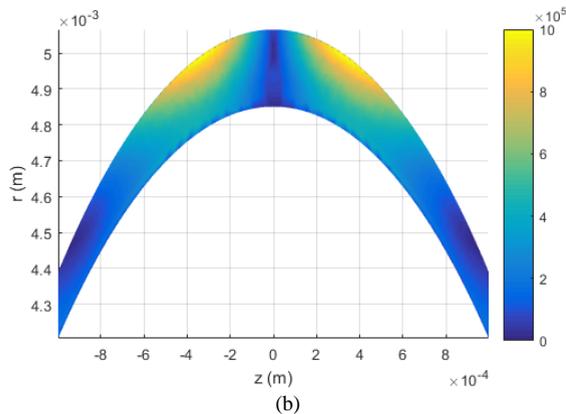

(b)

FIGURE 16. Eddy current distributions for more curved structures with different electrical conductivities and thickness under an operation frequency of 1 MHz a) The original aluminium curved structure with an electrical conductivity of 59.8 MS/m, and thickness of 20 μm b) The equivalent curved structure an electrical conductivity of 17.3 MS/m, and a thickness of 69 μm

## VI. CONCLUSION

For EM simulations with the FEM method, the sensor's response, i.e. mutual inductance is not easy to be computed especially under the high frequency. An extremely fine mesh is required to accurately simulate eddy current skin effects especially at high frequencies, and this could cause an extremely large total mesh for the modelling. In this paper, an equivalent-effect phenomenon is found, in which an alternative thicker structure but with less conductivity can produce the same impedance value as the original structure if a reciprocal relationship between the electrical conductivity and the thickness of the structure is observed. Since the equivalent structure has fewer mesh elements, the calculation burden can be significantly relieved when using the FEM method. The proposed equivalent-effect phenomenon has been validated from the measurements, analytical and FEM simulations for several types of structures.